\documentclass[a4paper,12pt]{article}
\usepackage[english]{babel}
\usepackage{times}
\usepackage{epsfig}
\usepackage{cite}

\def\slantfrac#1#2{\hbox{$\,^{#1}\!/_{#2}$}}
\def\mdot{{\raisebox{1pt}{\hbox{$\stackrel{\bullet}{M}$}}}\ \!\!}
\newcommand{\grad}{\mathop{\rm grad}\nolimits}
\newcommand{\mvec}[1]{{\bf #1}}
\newcommand{\ddiv}{\mathop{\rm div}\nolimits}
\def\dfrac#1#2{{\displaystyle#1\over\displaystyle#2}}
\newcommand{\Lone}{\mathop{\rm L_1}\nolimits}
\newcommand{\eqref}[1]{(\ref{#1})}

\title{Morphology of the Interaction Between the Stream and Cool Accretion
Disk in a Semi-detached Binary Systems}

\author{D.V.~Bisikalo$^1$, A.A.~Boyarchuk$^1$, P.V.~Kaygorodov$^1$,
O.A.~Kuznetsov$^{1,2}$\\[1cm]
\small{$^1$Institute of Astronomy, Russian Academy of Sciences,Moscow, Russia}\\
\small{$^2$Keldysh Institute of Applied Mathematics, Moscow, Russia}
}

\date{2003}

\begin{document}

\maketitle

\begin{abstract}
We analyze heating and cooling processes in accretion disks in
binaries. For realistic parameters of the accretion disks in close binaries
($\mdot\simeq10^{-12}\div10^{-7}M_\odot/year$ and
$\alpha\simeq10^{-1}\div10^{-2}$), the gas temperature in the outer
parts of the disk is $\sim{}10^4$~K to $\sim{}10^6$~K.
Our previous gas-dynamical studies
of mass transfer in close binaries indicate that, for hot disks (with
temperatures for the outer parts of the disk of several hundred thousand~K),
the interaction between the stream from the inner Lagrange point and the
disk is shockless. To study the morphology of the interaction between the
stream and a cool accretion disk, we carried out three-dimensional modeling
of the flow structure in a binary for the case when the gas temperature in
the outer parts of the forming disk does not exceed 13~600~K. The flow
pattern indicates that the interaction is again shockless. The computations
provide evidence that, as is the case for hot disks, the zone of enhanced
energy release (the ``hot line'') is located beyond the disk, and originates
due to the interaction between the circum-disk halo and the stream.
\end{abstract}

\section{Introduction}
\label{sec1}

In 1999-2002, we developed a three-dimensional, gas-dynamical model and used
it to study the flow patterns in binary 
systems~\cite{cit1,cit2,cit3,cit4,cit5,cit6,cit7,cit8,cit9,cit10,cit11,cit12,cit13,cit14}.
 These studies
indicate that the flow structure is substantially affected by rarefied gas
of the intercomponent envelope. In particular, a self-consistent solution
does not include a shock interaction between the stream from the inner
Lagrange point $\Lone$ and the forming accretion disk (a ``hot spot'').
The region
of enhanced energy release (the ``hot line'') is located beyond the disk and
is due to the interaction between the envelope and the stream. However,
these solutions were obtained for temperatures of the outer parts of the
accretion disk of 200~000 -- 500~000~K. To check if this behavior is
universal, the morphology of the flow must be considered for various disk
temperatures.

First, we will study here the interval of plausible temperatures of
accretion disks in close binaries. In Section~\ref{sec2}, based on an analysis of
heating and cooling in accretion disks, we will show that, for realistic
parameters of the disks in close binaries, 
($\mdot\simeq10^{-12}\div10^{-7}M_\odot/year$ and
$\alpha\simeq10^{-1}\div10^{-2}$), the gas temperature in the outer parts of
the disk is between 13~600~K
and $\sim 10^6$~K. This implies that cool accretion disks can form in some
close binaries.

Second, we will consider the morphology of the interaction
between
streams of matter and cool accretion disks in semi-detached binary systems
(Sections~\ref{sec3} and~\ref{sec4}). The basic problem here is whether the
interaction between the stream and the disk remains shockless, as was shown
for relatively hot disks \cite{cit1,cit3,cit4,cit8,cit14}.
Section~\ref{sec5} presents our main
conclusions and a physical basis for the universal nature of the shockless
interaction between the stream and disk.

\section{Heating and Cooling in Accretion Disks}
\label{sec2}

In this Section, we consider the temperature of an accretion disk for
various accretion rates, i.e., the dependence $T(\mdot)$.

\subsection{Basic Equations}
\label{sec21}

The vertical structure of an accretion disk is specified by the balance
between the vertical component of the gravitational force and the (vertical)
pressure gradient, which, in turn, is specified by the balance between
heating and cooling of the gas. The heating is associated with viscous
dissipation of kinetic energy, and also with bulk radiative heating, which,
in turn, is specified by the radiation of the central object. Cooling is
brought about by several mechanisms: bulk radiative cooling, radiative heat
conduction, and convection. Assuming that advective terms and terms
associated with adiabatic heating or cooling are small, the steady-state
energy equation
$$
Q^+-Q^-=0
$$
can be written as follows.

(1) For the optically thin case, when $Q^+$ is specified by bulk radiative
heating and viscous heating and $Q^-$ is determined by bulk radiative cooling,
\begin{equation}
Q^+_{visc}(\rho,T)+n^2\cdot(\Gamma(T,T_{wd})-\Lambda(T))=0\,.
\label{eq1a}
\end{equation}

Here, $\Gamma(T,T_{wd})$ is the radiative-heating function, which depends on
the
gas temperature $T$ and the temperature of the central object $T_{wd}$, 
$\Lambda(T)$ is the
radiative-cooling function, and $Q^+_{visc}(\rho,T)$ is the viscous heating.

(2) For the optically thick case, $Q^+$ is specified by viscous heating,
while $Q^-$
is specified by radiative heat conduction\footnote{We neglect molecular
heat conduction since it is very small compared with the radiative heat
conduction.} and convection in the vertical direction:
\begin{equation}
Q^+_{visc}(\rho,T) -\dfrac{\partial F_{rad}}{\partial z}
-\dfrac{\partial F_{conv}}{\partial z} =0\,.
\label{eq1b}
\end{equation}
Here, $F_{rad}$ and $F_{conv}$ are the radiative and convective energy fluxes.
To determine the functions in~\eqref{eq1a} and~\eqref{eq1b}, we will need 

\begin{itemize}

\item[--] the equation of continuity
\setcounter{equation}{1}
\begin{equation}
-\mdot=2\pi\int r\cdot\rho\cdot v_r {~\rm d}z={\rm const}\,,
\label{cont}
\end{equation}

\item[--] the equation of angular-momentum balance $\lambda\equiv
r^2\Omega_K$ in the radial direction:
\begin{equation}
\dfrac{\partial}{\partial r} \left(r\cdot\rho\cdot v_r\cdot\lambda
\right)= \dfrac{\partial}{\partial r} \left(\nu\cdot\rho\cdot
r^2\cdot \dfrac{\partial\Omega_K}{\partial r}\right)\,,
 \label{angmom}
\end{equation}
from which it follows that
\begin{equation}
|v_r|=-\nu\cdot\Omega_K'\cdot\Omega^{-1}_K\cdot r^{-1}
\simeq\nu\cdot r^{-1}\,,
 \label{angmom1}
\end{equation}

\item[--] and the viscous heating
\begin{equation}
Q^+_{visc}=\rho\cdot\nu\cdot\left(r\cdot\dfrac{\partial\Omega_K}{\partial
r}\right)^2\,.  
\label{vischeat}
\end{equation}

\end{itemize}

Here, $\mdot$ is the accretion rate, $\Omega_K=\sqrt{GM/r^3}$
the angular velocity of the
Keplerian rotation of the disk, $M$ the mass of the central object, $G$ the
gravitational constant, $\rho$ -- the density, $v_r$ the radial velocity, and
the $\nu$ -- coefficient of kinematic viscosity. 
Note that the molecular viscosity cannot
provide the necessary dissipation, and dissipation processes are usually
considered to be associated with turbulent or magnetic viscosity.

To determine the vertical pressure gradient, we will use the equation of
hydrostatic balance in the vertical direction
\begin{equation}
\dfrac1\rho\cdot \dfrac{\partial P}{\partial z}=
\dfrac{\partial}{\partial z} \left(\dfrac{GM}{\sqrt{r^2+z^2}}\right)
\simeq -\Omega_K^2 z\,, \label{vert}
\end{equation}
as well as the equation of state of an ideal gas with radiation
$$
P=\rho{\cal R}T+\slantfrac13aT^4\,.
$$
Here, $P$ is the pressure, $T$ the temperature, ${\cal R}$ the gas constant,
and $a$ the radiation constant. All equations are given in cylindrical
coordinates, $(r,z)$.

\subsection{The Solution Method}
\label{sec22}

To determine the dependence $T(\mdot)$, we will use \eqref{cont} and
\eqref{angmom1} together with
the expression for the viscosity coefficient . We will use the formula for
suggested by Shakura~\cite{cit15}, $\nu=\alpha c_s H$, where $H$ is the height of the
disk and $c_s\simeq\sqrt{{\cal R}T+\slantfrac13aT^4/\rho}$
is the sound speed. If we neglect the $z$ dependence of the
density and use $\bar\rho$ averaged over the height (further, we will denote
this
quantity simply as $\rho$), the integration of \eqref{vert}
yields the height of the
disk $H$:
$$
H=c_s\cdot\Omega_K^{-1}\,.   
$$
We will determine $c_s$ from the temperature in the equatorial plane of the
disk, $z=0$. This approach is sufficiently correct for our purposes due to the
uncertainty in the parameter $\alpha$. As a result, we obtain an equation 
relating $\mdot$,
$\left.T\vphantom{x^x_x}\right|_{z=0}$, and $\rho$ for the specified
$r$ and $\alpha$,
\begin{equation}
\mdot=2\pi\cdot\alpha\cdot\Omega_K^{-2}\cdot\rho\cdot  
c_s^3=2\pi\cdot\alpha\cdot\Omega_K^{-2}\cdot\left({\cal  
R}T\rho^{2/3}+\slantfrac13aT^4\rho^{-1/3}\right)^{3/2}\,.
\label{appeq1}
\end{equation}
This equation reduces to a cubic equation in the variable $\rho^{1/3}$, and
its solution has two branches: one with a negative real root and two complex
ones, and one with three real roots, one of which is negative. Only positive
real roots for the density are physically meaningful. For such roots to
exist, the following condition must be satisfied:
\begin{equation}
\mdot>\dfrac{\sqrt{3}\pi\cdot\sqrt{\cal R}\cdot a\cdot\alpha \cdot
T^{9/2}}{\Omega_K^2}\,.
\label{estim1}
\end{equation}
which yields the minimum accretion rate for the given $T$, $r$, and $\alpha$.
When
deriving this condition, we used the equation of state taking into account
the radiation pressure.

This estimate can also be written in the form
$$
T<7\cdot10^5\left(\dfrac r{R_{wd}}\right)^{-2/3}
\left(\dfrac{\mdot}{10^{-9}M_\odot/year}\right)^{2/9}
\left(\dfrac\alpha{0.1}\right)^{-2/9}\quad {\rm K}\,,
$$
where $R_{wd}=10^9cm$ is the radius of the accretor (white dwarf).

Let us consider the condition \eqref{estim1} for the outer parts of the accretion disk.
Let us take $r=A/5$, where $A$ is the distance between the components of the
binary $A=1.42R_\odot$), which corresponds to situation for the dwarf nova 
IP~Peg; as a result, we obtain
\begin{equation}
\mdot>10^{-9}\left(\dfrac{T}{10^5~{\rm K}}\right)^{9/2}
\left(\dfrac\alpha{0.1}\right)\quad M_\odot/year\,.
\label{estim2}
\end{equation}
If \eqref{estim1} is satisfied, the roots of Eq.~\eqref{appeq1} relating $\rho$, $T$, and $\mdot$
for a given $r$ and $\alpha$ can be written
$$
\rho=({\cal R}T)^{-3/4}
\left(\dfrac{\mdot\Omega_K^2}{2\pi\alpha}\right)
\sin^3\left(\slantfrac13 \arcsin\left(\sqrt{\cal R}aT^{9/2}
\dfrac{2\pi\alpha}{\mdot\Omega_K^2}\right) \right)\,,
$$
  
$$
\rho=({\cal R}T)^{-3/4}
\left(\dfrac{\mdot\Omega_K^2}{2\pi\alpha}\right)
\cos^3\left(\slantfrac13 \arcsin\left(\sqrt{\cal R}aT^{9/2}
\dfrac{2\pi\alpha}{\mdot\Omega_K^2}\right) +\dfrac\pi6\right)
$$
(for simplicity, we have omitted numerical factors
$\slantfrac{\sqrt{3}}2\simeq1$). The first of
these corresponds to disks with dominant radiation pressure 
($\beta=\slantfrac13aT^4/\rho{\cal R}T>1$)
and the second to disks with dominant gas pressure ($\beta<1$).

These formulas describe the two branches of the two-parameter dependence 
$\rho(\mdot,T)$. To calculate the dependence $T(\mdot)$, we must use the
additional heat balance equations~\eqref{eq1a}--\eqref{eq1b}.
As follows from Section~\ref{sec21}, the form
of~\eqref{eq1a}--\eqref{eq1b} depends on the optical depth of the disk, which, accordingly, must
be calculated.

\subsection{Optical Depth}
\label{sec23}

The optical depth $\tau$ is specified by the product of the absorption
coefficient $\kappa$, the density, and the geometrical depth of the layer
$l$: $\tau=\kappa\cdot\rho\cdot l$. For disk
accretion, the basic parameter is the ratio of the geometrical depth of the
layer where $\tau=1$ and the height of the disk: $l^{\tau=1}/H$.
After simple manipulation, we obtain
\begin{equation}
\dfrac{l^{\tau=1}}{H}
=\dfrac{2\pi\cdot\alpha}{\kappa\cdot\mdot}\cdot
c_s^2\cdot\Omega_K^{-1}\,.
\label{eq4}
\end{equation}

\begin{figure}[t]
\centerline{\hbox{\epsfig{figure=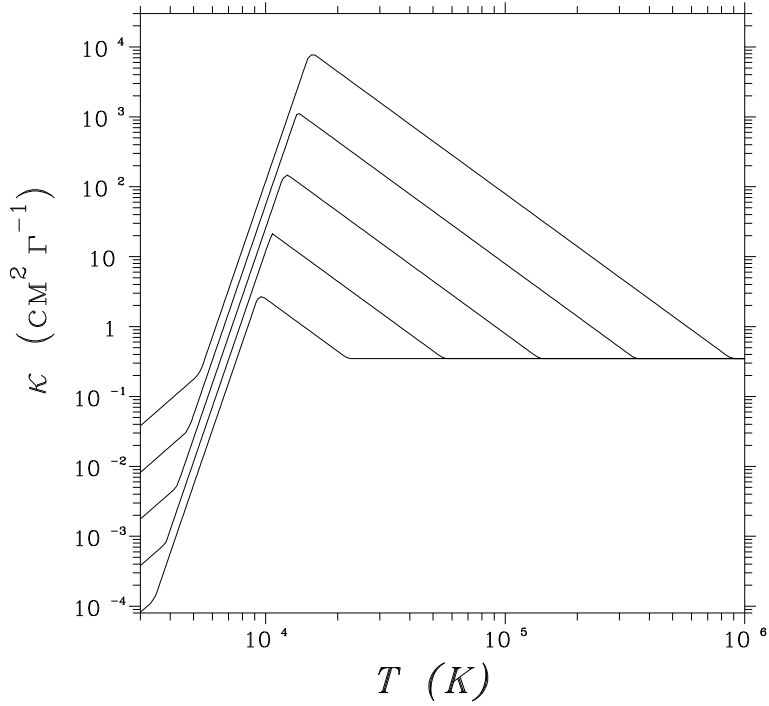,width=10cm}}}
\caption{\small The $\kappa(T)$ dependence for $n=10^{18}cm^{-3}$, 
$n=10^{17}cm^{-3}$, $n=10^{16}cm^{-3}$, $n=10^{15}cm^{-3}$, and 
$n=10^{14}cm^{-3}$ (top to bottom)~\protect\cite{cit18}.}
\label{fig1}
\end{figure}

The absorption coefficient $\kappa$ displays a complicated dependence on $T$
and $\rho$ (and
also on the degree of ionization, chemical composition, etc.). Here, we
adopted the simple approximation for $\kappa(T,\rho)$~\cite{cit16,cit17,cit18}
$$
\kappa(T,\rho)=\left\{
\begin{array}{ccc}
\kappa_1\cdot\rho^{2/3}\cdot T^{3}\,,
         &\qquad \kappa_1=10^{-8}\,, 
         &\qquad (\kappa1)\\
~\\
\kappa_2\cdot\rho^{1/3}\cdot T^{10}\,,
         &\qquad \kappa_2=10^{-36}\,, 
         &\qquad (\kappa2)\\
~\\
\kappa_3\cdot\rho\cdot T^{-5/2}\,,
         &\qquad \kappa_3=1.5\cdot10^{20}\,,
         &\qquad (\kappa3)\\
~\\
\kappa_4\,,
         &\qquad \kappa_4=0.348\,.
         &\qquad (\kappa4)
\end{array}\right.
$$
According to~\cite{cit18}, these four subregions correspond to scattering on
molecular hydrogen, scattering on atomic hydrogen, free--free and free--bound
transitions, and Thompson scattering. The boundaries of the sub-regions,
i.e. the transitions from one expression to another, are specified by the
equality of the $\kappa$ values calculated from these expressions.
Figure~\ref{fig1}
presents the dependences of $\kappa$ on $T$ and $\rho$. We can see regions
with $d\kappa/dT>0$, where
thermal instability can develop when the dependence between the surface
density and the disk temperature forms an S curve in the $(\Sigma,T_{eff})$
plane. Thermal instability is often invoked to explain the phenomenon of
dwarf novae (see, for example,~\cite{cit19,cit20});
however, it is clear that this can
occur only for sufficiently cool disks.

\begin{figure}[t]
\centerline{\hbox{\epsfig{figure=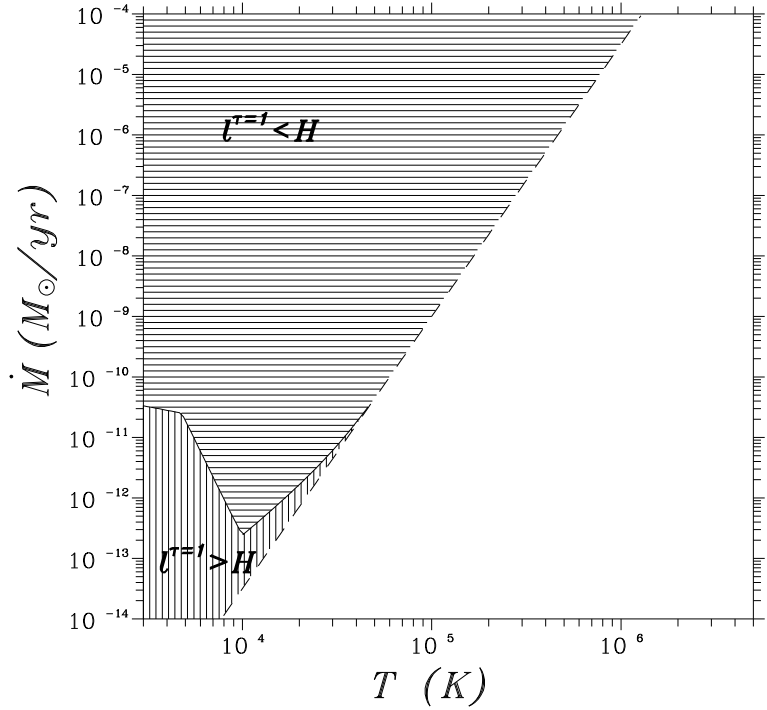,width=10cm}}}
\caption{\small Solution of~\eqref{appeq1} for disks with dominant gas pressure in
the $(T,\protect\mdot)$
plane for $\alpha=0.1$ and $r=A/5$. The horizontal thick shading indicates
optically thick disks, and vertical shading optically thin disks; the solid
line marks the border between these regions. The dashed line corresponds to
condition~\eqref{estim2} for the existence of the solution of~\eqref{appeq1};
there is no solution
below this line.}
\label{fig2a}
\end{figure}

\begin{figure}[t]
\centerline{\hbox{\epsfig{figure=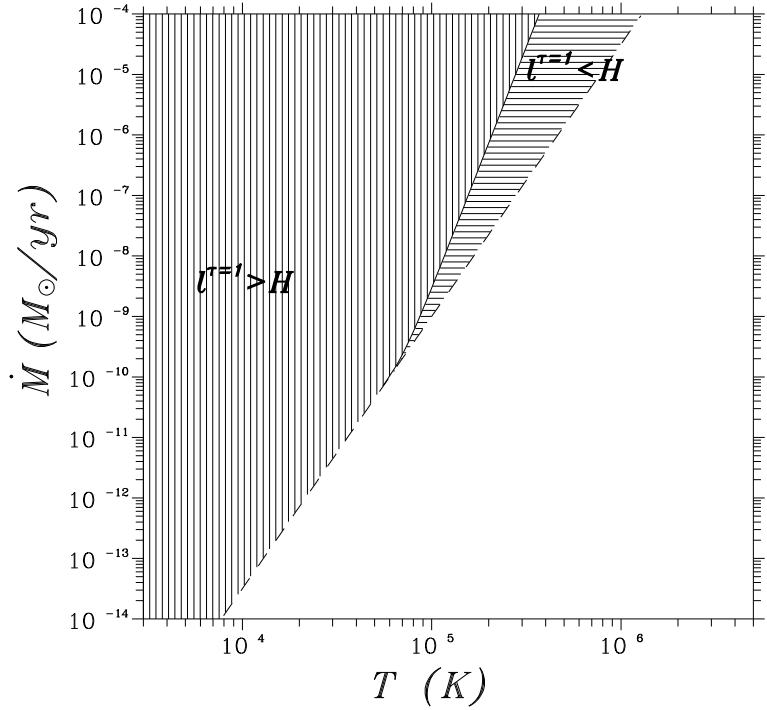,width=10cm}}}
\caption{\small The same as Fig.~\ref{fig2a} for disks with dominant radiation
pressure.}
\label{fig2b}
\end{figure}

Let us return to~\eqref{appeq1}, taking $\alpha=0.1$ and $r=A/5$ and considering disks
with dominant gas pressure, for which $\beta=1/3aT^4/\rho{\cal R}T<1$.
The shaded region in the $(T,\mdot)$ plane in Fig.~\ref{fig2a} corresponds to all
possible solutions for these
disks. The dashed line corresponds to condition~\eqref{estim2} for the existence of a
solution for~\eqref{appeq1} -- there is no solution below this line. The solid line
indicates the boundary between the optically thick and optically thin
solutions: the horizontal shading marks the region of optically thick disks,
while the vertical shading marks optically thin disks. Figure~\ref{fig2b} presents a
similar pattern for disks with dominant radiation pressure ($\beta>1$).

We can see from Fig.~\ref{fig2a} that, for realistic values
$\mdot\in[10^{-12},10^{-7}] M_\odot/year$,
disks with dominant gas pressure are mainly optically thick, though
solutions corresponding to optically thin cool disks are possible for small
$\mdot$. It follows from Fig.~\ref{fig2b} that disks with dominant radiation pressure
are mainly optically thin; optically thick hot disks can exist only for high
$\mdot$.

\subsection{Optically Thick Disks}
\label{sec24}

In Section~\ref{sec22}, we derived Eq.~\eqref{appeq1}, which relates $\mdot$, $T$, and $\rho$
for given $a$, $r$ and $\alpha$ . Using the supplementary heat-balance
equations~\eqref{eq1a}--\eqref{eq1b}, we can reduce the number of unknowns and obtain the desired
relation between $\mdot$ and $T$.

The vertical temperature distributions in optically thick disks are
described by the equation of radiative heat conduction with a source due to
viscous heating~\eqref{eq1b}, which can be written in the form
\begin{equation}
\dfrac{\partial e}{\partial t}= \dfrac{\partial}{\partial
z}\left(\dfrac{1}{\kappa\rho} \dfrac{\partial}{\partial  
z}\left(\slantfrac{1}{3}acT^4\right)\right) +\rho\alpha
c_s^2\Omega_K\,, \label{difur}
\end{equation}
where $e$ is the specific internal energy and $c$ is the velocity of light.
To solve~\eqref{difur}, we must specify boundary conditions. Due to the symmetry of
the problem, the temperature derivative in the equatorial plane must be zero;
i.e., $\left.T'\vphantom{x^x_x}\right|_{z=0}=0$. The temperature at the upper
boundary of the disk is specified by the condition
$\Gamma(T_*,T_{wd})=\Lambda(T_*)$. Though the functions $\Gamma(T,T_{wd})$ and
$\Lambda(T)$ are complex, they are known and can be found in the literature
(see, for example,~\cite{cit21,cit22,cit23}).
The temperature derived by equating these
functions (for a temperature of the central object (white dwarf) of 
$T_{wd}=70~000$~K) is $T(H)=T_*=13~600$~K.

The solution of~\eqref{difur} enters a stationary regime when the characteristic
heat-conduction time
$$
t_{diff}\simeq\dfrac{{\cal R}\kappa\rho^2H^2}{acT^3}
$$
is comparable to the time for viscous heating
$$
t_{heat}\simeq\dfrac{{\cal R}T}{\alpha c_s^2\Omega_K}\simeq
\alpha^{-1}\Omega_K^{-1}\,.
$$
Note that~\eqref{difur} can be integrated analytically in the steady-state case. Let
us denote $U=T^4$, $U_*=T_*^4$,
$U_0=\left.U\vphantom{x^x_x}\right|_{z=0}$ and again assume that does not
depend on $z$. Then,
$$
\dfrac{d}{d z}\left(\dfrac{1}{\kappa\rho} \dfrac{d}{d
z}\left(\dfrac{ac}{3}U\right)\right) =-\rho\alpha c_s^2\Omega_K\,.
$$
After integrating over $z$, we obtain
$$
\dfrac{1}{\kappa\rho} \dfrac{d}{d z}\left(\dfrac{ac}{3}U\right)
=-\rho\alpha c_s^2\Omega_K z\,,
$$
The integration constant is equal to zero, since
$\left.U'\vphantom{x^x_x}\right|_{z=0}=0$. For
convenience, we will transform this last equation to the form
$$
\dfrac{1}{\kappa} \dfrac{dU}{d z}\equiv\dfrac{dB}{d z}
=-\dfrac{3}{ac}\rho^2\alpha c_s^2\Omega_K z\,,
$$
where the function $B(U)$ is determined from the differential equation 
$\displaystyle\dfrac{dB}{dU}=\dfrac{1}{\kappa(U,\rho)}$ and can be written
in an analytical form if $\rho$ is fixed. Integrating this last equation
over $z$, we obtain
$$
B(U)=B(U_*)+\dfrac{3}{2ac}\rho^2\alpha c_s^2\Omega_K(H^2-z^2)\,,
$$
or, for $z=0$
$$
B(U_0)=B(U_*)+\dfrac{3}{2ac}\rho^2\alpha c_s^2\Omega_KH^2\,.
$$
Using the expressions
$$
c_s^2= \left({\cal
R}U_0^{1/4}+\slantfrac13\dfrac{aU_0}{\rho}\right)\,,
$$
$$
H^2= \left({\cal R}U_0^{1/4}+\slantfrac13\dfrac{aU_0}{\rho}\right)
\Omega_K^{-2}
$$
we obtain the algebraic equation for $U_0$
$$
B(U_0)=B(U_*)+\dfrac{3}{2ac}\rho^2\alpha \Omega_K^{-1} \left({\cal
R}U_0^{1/4}+\slantfrac13\dfrac{aU_0}{\rho}\right)^2\,.
$$
This equation implicitly specifies the dependence $U_0(\rho)$; i.e., $T(\rho)$.
Expressing $\mdot$ in terms of $\rho$ and $T$, we can derive the dependence 
$\mdot(\rho)=\mdot(T(\rho),\rho)$, which
yields the dependence $T(\mdot)$ in parametric form. Formally, the resulting
solution can also exist in optically thin regions; however, given the
adopted assumptions, these points can be rejected.

\begin{figure}[t]
\centerline{\hbox{\epsfig{figure=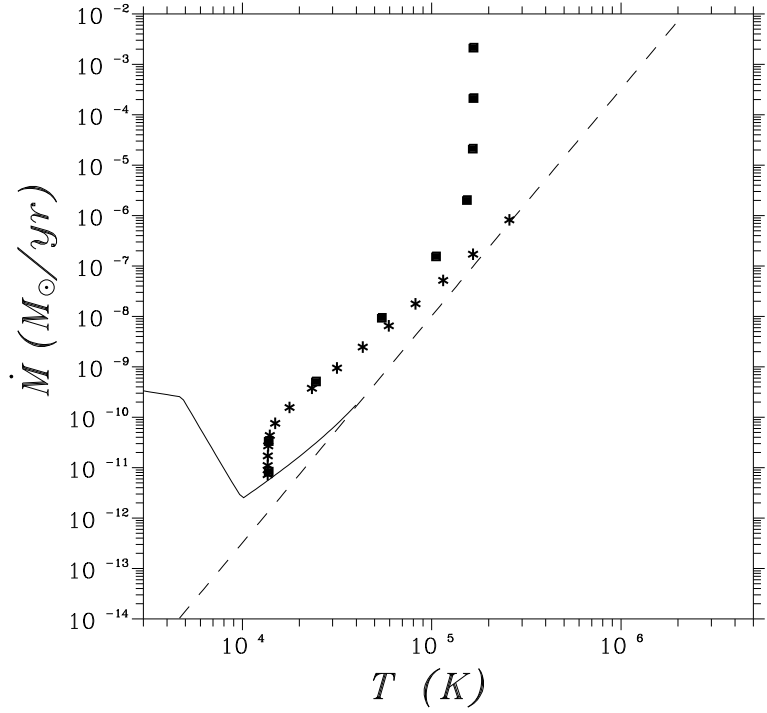,width=10cm}}}  
\caption{\small Solution of~\eqref{difur} for an optically thick disk for
$\alpha=1$ and $r=A/5$ (asterisks). Solutions of~\eqref{difur1} taking into account
convection are labeled by squares. The dashed line is a lower bound for the
region in which there exist solutions of~\eqref{appeq1}, and the solid line separates
the regions of optically thin and optically thick disks.}
\label{fig3a}

\end{figure}
\begin{figure}
\centerline{\hbox{\epsfig{figure=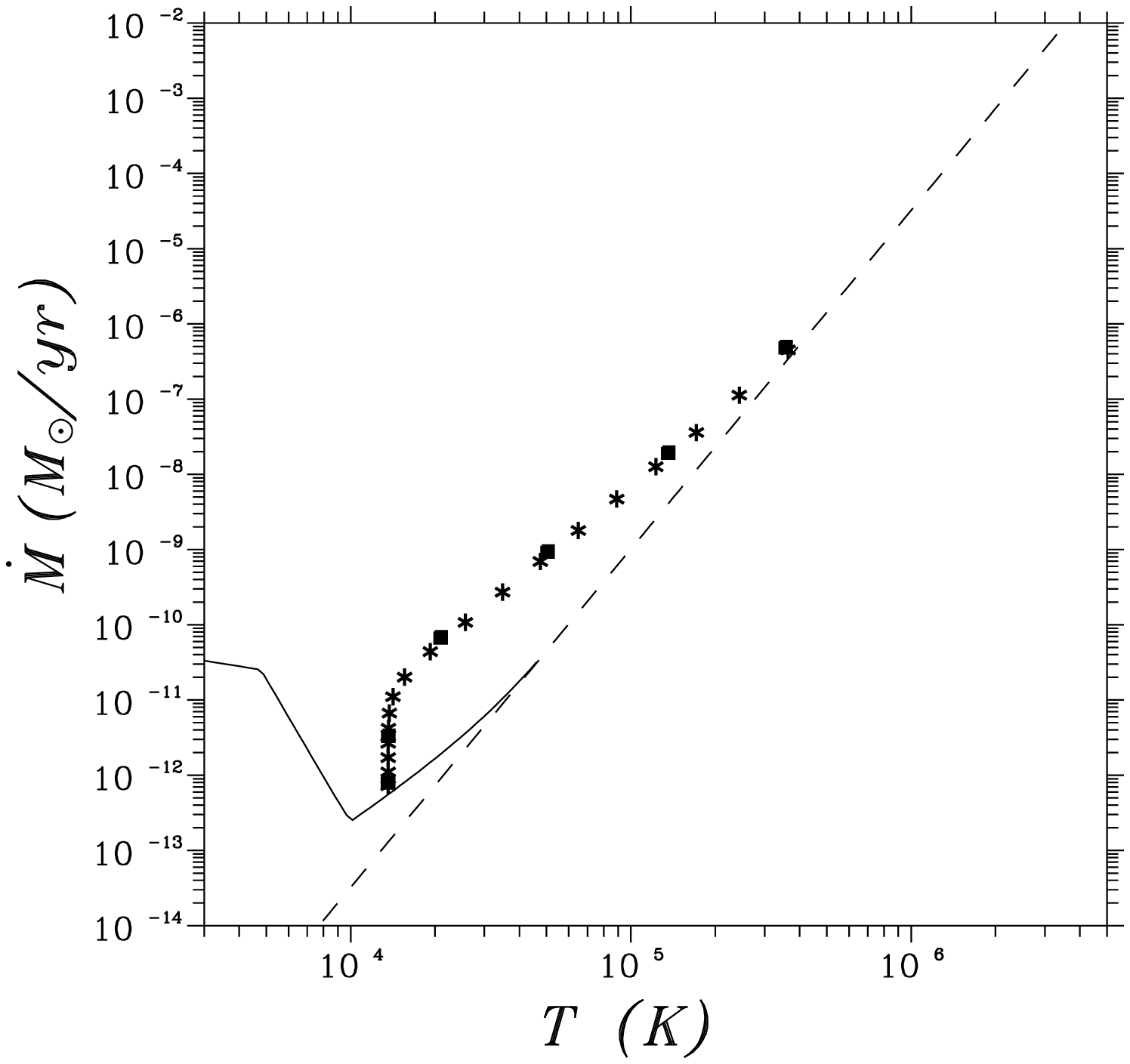,width=10cm}}}
\caption{\small The same as Fig.~\ref{fig3a} for $\alpha=0.1$.}
\label{fig3b}
\end{figure}

\begin{figure}[t]
\centerline{\hbox{\epsfig{figure=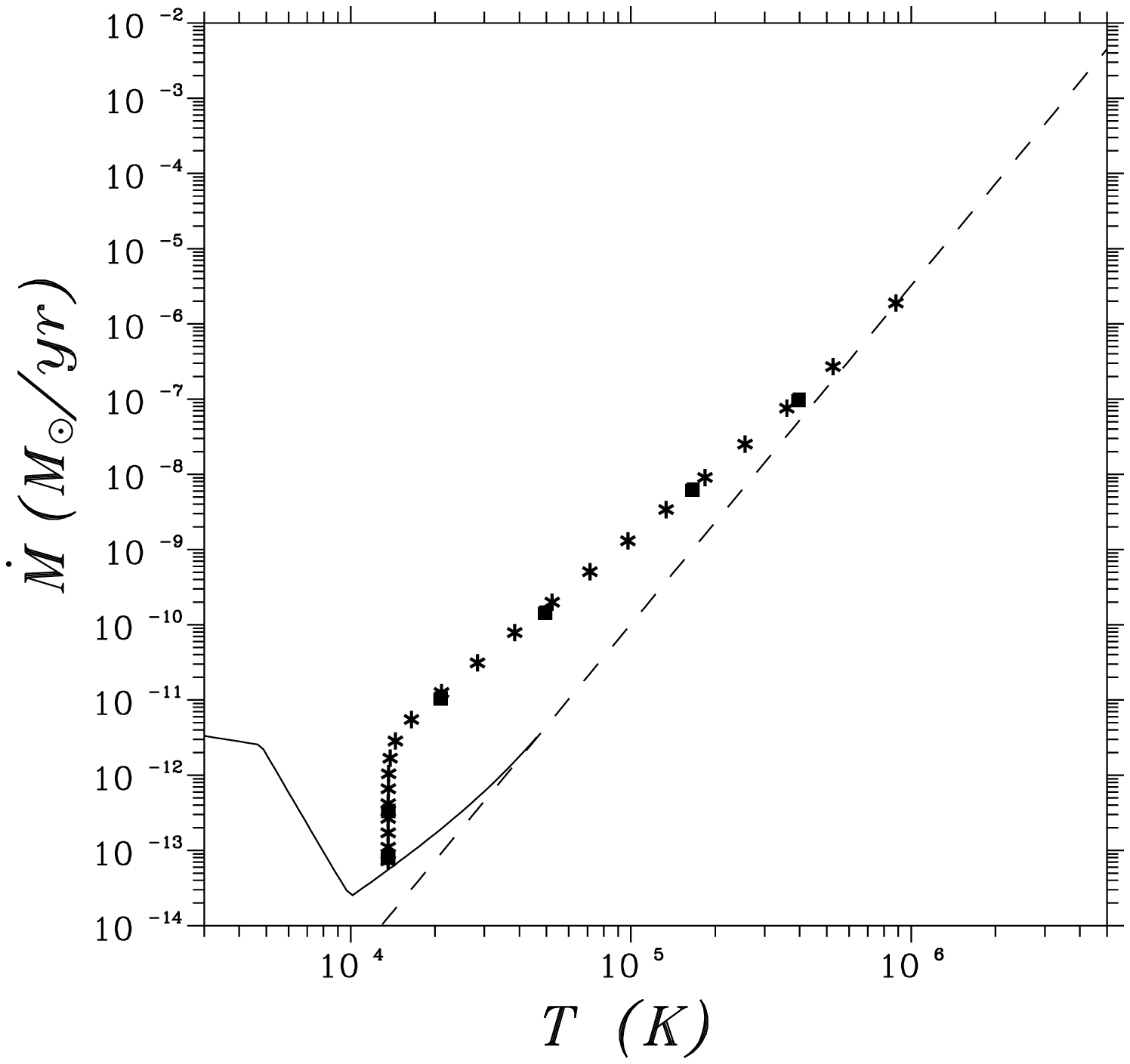,width=10cm}}}
\caption{\small The same as Fig.~\ref{fig3a} for $\alpha=0.01$.}
\label{fig3c}
\end{figure}

\begin{figure}[t]
\centerline{\hbox{\epsfig{figure=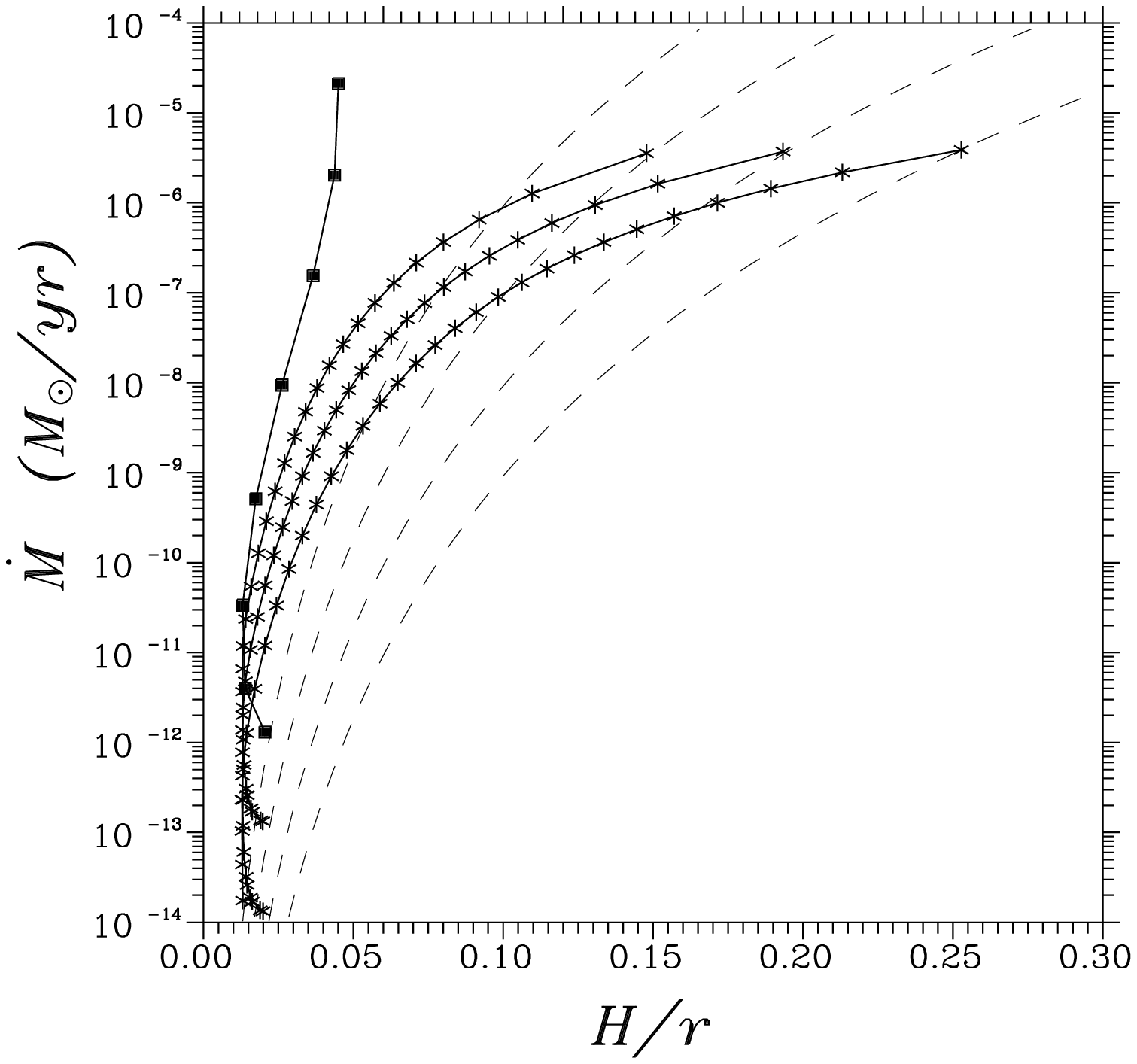,width=10cm}}}
\caption{\small A disk with dominant gas pressure. The solid
lines indicate possible states of the disk in the $H/r-\protect\mdot$
plane for $r=A/5$.
Solutions of~\eqref{difur} taking into account radiative heat conduction and viscous
heating are shown by the lines with asterisks for $\alpha=1$,
$\alpha=0.1$, $\alpha=10^{-2}$, $\alpha=10^{-3}$ (top to the bottom).
Solutions of~\eqref{difur1} taking into account radiative heat
conduction, convection, and viscous heating are shown by the lines with
squares for $\alpha=1$. The dashed lines bound from below regions in which the
solution of~\eqref{appeq1} can exist for $\alpha=1$,
$\alpha=0.1$, $\alpha=10^{-2}$, $\alpha=10^{-3}$ (top to bottom).}
\label{fig3d}
\end{figure}

Let us consider a graphical representation of the solution derived. 
Figure~\ref{fig3a} presents the dependence $T(\mdot)$ for $\alpha=1$ and $r=A/5$,
marked by asterisks.
As in Fig.~\ref{fig2a}--~\ref{fig2b},
the dashed lines bound from below the domain in which there
exist solutions of~\eqref{appeq1}, while the solid line separates the domains of
optically thin and optically thick disks. Figures~\ref{fig3b}--\ref{fig3c}
display the solutions
for $\alpha=0.1$ and $\alpha=0.01$, respectively.
Figure~\ref{fig3c} presents the accretion rate
as a function of the disk thickness. We can see that all the obtained disks
are geometrically thin; i.e., $H\ll r$.

Radiative heat conduction is not the only mechanism for heat transfer into
optically thin regions. Under certain conditions, convection can also play a
substantial role. Neglecting the radiation pressure, the convective flux can
be written in the form~\cite{cit24,cit25}
$$
F_{conv}=c_P\cdot\rho\cdot\left(\dfrac{|g|}{T}\right)^{1/2}
\cdot\dfrac{l^2}{4} \cdot(\Delta\nabla T)^{3/2}\,,
$$
$$
\Delta\nabla T=-\dfrac{T}{c_P}\cdot \dfrac{\partial S}{\partial
z}\,.
$$   
Here, $c_P$ is the heat capacity at constant pressure, $S={\cal
R}\cdot\ln(T^{3/2}/\rho)$ the specific
entropy, $g=-\Omega_K^2 z$ the gravitational acceleration, and $l$
the mixing
length, taken to be $l=\alpha H$. To determine the vertical temperature
distribution taking convection into account,
we must solve the equation
\begin{equation}
\dfrac{\partial e}{\partial t}= \dfrac{\partial}{\partial
z}\left(\dfrac{1}{\kappa\rho} \dfrac{\partial}{\partial  
z}\left(\slantfrac{1}{3}acT^4\right)\right) -\dfrac{\partial
F_{conv}}{\partial z} +\rho\alpha c_s^2\Omega_K \label{difur1}
\end{equation}
with the same boundary conditions as for~\eqref{difur}.
Equation~\eqref{difur1} does not admit
a simple analytical solution, and we solved this equation numerically using
the method of establishment. The solution is denoted by the squares in 
Figs.\ref{fig3a}--\ref{fig3c}.
We can see that convection plays a significant role only when
$\alpha\simeq1$.

Summarizing, we can assert that, in the optically thick case with small 
$\mdot$, the disk displays the constant temperature $T=T_*=13600^\circ$~K,
while the temperature increases as $T\propto\mdot^{1/3}$ at larger values of
$\mdot$. Thus, for
realistic parameters of the accretion disks in close binaries,
$\mdot\simeq10^{-12}\div10^{-7}M_\odot/year$ and
$\alpha\simeq10^{-1}\div10^{-2}$, the gas temperature in the outer parts of
the disk ($r\simeq A/5\div A/10$) is $10^4$~K to $\sim10^6$~K.

Solving~\eqref{difur} for various $r$, we can also calculate the dependences $T(r)$
and $\rho(r)$. The calculations indicate that $T\propto r^{-0.8}$ and
$\rho\propto r^{-1.8}$, which is
consistent with the dependence $T\propto r^{-3/4}$ obtained by Shakura and
Sunyaev~\cite{cit26}.

\subsection{Optically Thin Disks}
\label{sec25}

In this case, the temperature of the disk is specified by the balance between
radiative heating $\Gamma(T,T_{wd})$ and viscous heating~\eqref{vischeat}, on the one hand,
and radiative cooling $\Lambda(T)$, on the other. The heat-balance
equation~\eqref{eq1a} can be written
$$
\rho\alpha c_s^2\Omega_K+ \rho^2\cdot
m_p^2\cdot(\Gamma(T,T_{wd})-\Lambda(T))=0\,,
$$
which can be reduced to the quadratic equations in $\rho$
$$
\alpha\cdot(\rho{\cal R}T+\slantfrac13aT^4)\cdot\Omega_K+
\rho^2\cdot m_p^{-2}\cdot(\Gamma(T,T_{wd})-\Lambda(T))=0\,.
$$
The solution of this equation for specified $r$ and $\alpha$ yields the
dependence $\rho(T)$, and thus $T(\mdot)$. Formally,
this solution can also exist in optically thick regions, however these points
were rejected by virtue of the adopted assumptions.

It is shown in Section~\ref{sec23} that disks in which gas pressure dominates are
primarily optically thick, and solutions that correspond to optically thin
disks are possible only for small $\mdot$. Disks in which radiation pressure
dominates are primarily optically thin. The domination of radiation pressure
is possible only in the inner parts of the disk; therefore, we will adopt 
$r=A/20$ for the further analysis. For the typical dwarf nova IP~Peg, this
corresponds to five radii of the accretor (white dwarf).

\begin{figure}
\centerline{\hbox{\epsfig{figure=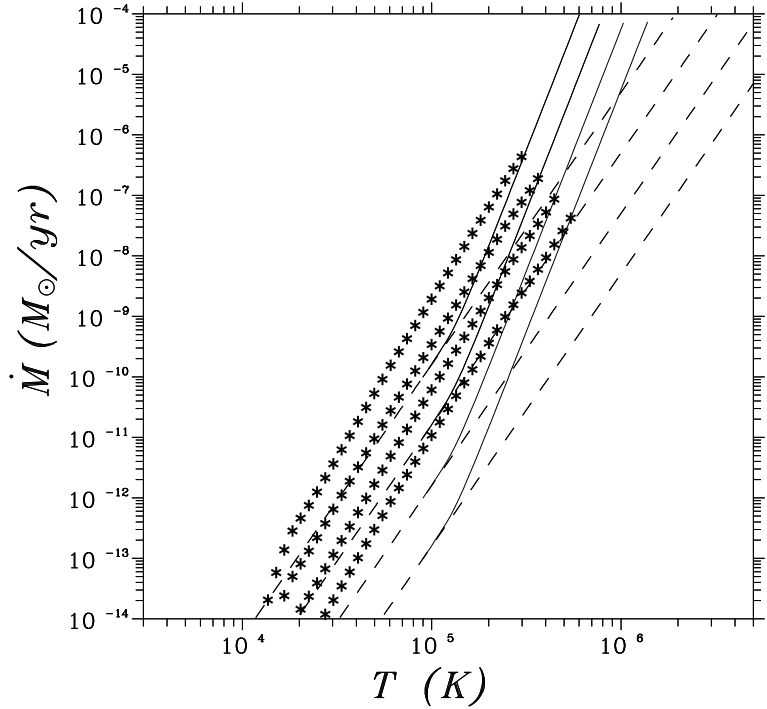,width=10cm}}}
\caption{\small Solutions for an optically thin disk for $\alpha=1$,
$\alpha=0.1$, $\alpha=10^{-2}$,
$\alpha=10^{-3}$ (top to bottom) and $r=A/20$ (asterisks).
The dashed lines bound
from below the domain in which there exists a solution of~\eqref{appeq1}; the solid lines
separate the regions for optically thin and optically thick disks.
}
\label{fig4}
\end{figure}

Figure~\ref{fig4} presents the results of our calculations; the asterisks denote the
$T(\mdot)$ dependences for $\alpha=1$,
$\alpha=0.1$, $\alpha=10^{-2}$,
$\alpha=10^{-3}$ (top to bottom), and $r=A/20$.
The disks obtained in these solutions are geometrically thick, 
$H\simeq r$. Note that the initial assumptions of the model restrict its
applicability: it is suitable only for geometrically thin disks, and the
solutions for geometrically thick disks are purely formal.

\section{The Model}
\label{sec3}

We described the flow structure in a binary system using a system of
gravitational gas-dynamical equations taking into account radiative heating
and cooling of the gas for the optically thin case:
\begin{equation}
\left\{
\begin{array}{l}
\dfrac{\partial\rho}{\partial t}+\ddiv\rho{\mvec{v}}=0\,,\\[5mm]
\dfrac{\partial\rho\mvec{v}}{\partial t}
+\ddiv(\rho\mvec{v}\otimes\mvec{v})+\grad P=-\rho\grad\Phi\,,\\[5mm]
\dfrac{\partial\rho(\varepsilon+|\mvec{v}|^2/2)}{\partial t}
+\ddiv\rho\mvec{v}(\varepsilon+P/\rho+|\mvec{v}|^2/2)=\\[5mm]
\qquad\qquad
=-\rho\mvec{v}\grad\Phi+\rho^2m_p^{-2}\left(\Gamma(T,T_{wd})-\Lambda(T)\right).
\end{array}
\right. \label{HDC}
\end{equation}
Here, as usual, $\rho$ is the density, $\mvec{v}=(u,v,w)$ the velocity vector,
$P$ the pressure, $\varepsilon$ the internal energy, $\Phi$ the Roche
gravitational potential, $m_p$ the
proton mass, and  $\Gamma(T,T_{wd})$ and $\Lambda(T)$ the radiative heating
and cooling functions, respectively. The system of gas-dynamical equations was
closed with the Clapeyron equation $P=(\gamma-1)\rho\varepsilon$, where 
$\gamma$ is the adiabatic index. We took the parameter $\gamma$ to be 5/3.

Our main goal here is to study the morphology of the interaction between the
stream and the cool accretion disk. It follows from Section~\ref{sec2} that the outer
parts of the accretion disk can be cool for small $\mdot$ and, in particular,
in
the case of an optically thin disk. The system of equations~\eqref{HDC} enables us
to carry out three-dimensional modeling of the flow structure in a binary
within our formulation of the problem. In the model, the temperature of the
disk is 13~600~K.

We solved this system of equations using the Roe-Osher
method~\cite{cit14,cit27,cit28},
adapted for multiprocessing computations via spatial decomposition the
computation grid (i.e., partitioning into subregions, with synchronization
of the boundary conditions)~\cite{cit29}. We considered a semi-detached binary
system containing a donor with mass $M_2$ filling Roche lobe and an accretor
with mass $M_1$. The system parameters were specified to be those of the dwarf
nova IP~Peg: $M_1=1.02M_\odot, M_2=0.5M_\odot, A=1.42R_\odot$.

The modeling was carried out in a non-inertial reference frame rotating with
the binary, in Cartesian coordinates in a rectangular three-dimensional
grid. Since the problem is symmetrical about the equatorial plane, only half
the space occupied by the disk was modeled. To join the solutions, we
specified a corresponding boundary condition at the lower boundary of the
computation domain. The accretor had the form of a sphere with radius
$10^{-2}A$.
All matter that ended up within any of the cells forming the accretor was
taken to fall onto the star. A free boundary condition was specified at the
outer boundaries of the disk -- the density was constant 
($\rho_b=10^-8\rho_{\Lone}$), where $\rho_{\Lone}$ is the density at the 
point $\Lone$, the temperature was 13~600~K, and the
velocity was equal to zero. The stream was specified in the form of a
boundary condition: matter with temperature 5800~K, density 
$\rho_{\Lone}=1.6\times 10^{-8} g/cm^3$ and velocity along the $x$ axis 
$v_x=6.3 km/s$ was injected into a zone
around $\Lone$ with radius $0.014A$. 
For this rate of matter input into the system, the model accretion rate was 
$\sim 10^{-12} M_{\odot}/year$.

The size of the computation domain, $1.12A\times 1.14A\times  0.17A$, was
selected so that it entirely contains both the disk and stream, 
including the point $\Lone$.
The computation grid with $121\times 121\times 62$ cells was distributed
between 81 processors, which constituted a two-dimensional $9\times 9$ matrix.

To increase the accuracy of the solution, the grid was made denser in the
zone of interaction between the stream and disk, making it possible to
resolve well the formed shock wave. The grid was also denser towards the
equatorial plane, so that the vertical structure was resolved, even for such
a cool disk.

We used the solution obtained for a model without cooling as the initial
conditions~\cite{cit12}. The model with cooling was computed during approximately
five revolutions of the system, until the solution became established. The
total time of the computations was $\approx$ 1000~h on the MBC1000A 
computer of the Joint Supercomputer Center (JSC).

\section{Computation Results}
\label{sec4}

\begin{figure}
\centerline{\hbox{\epsfig{figure=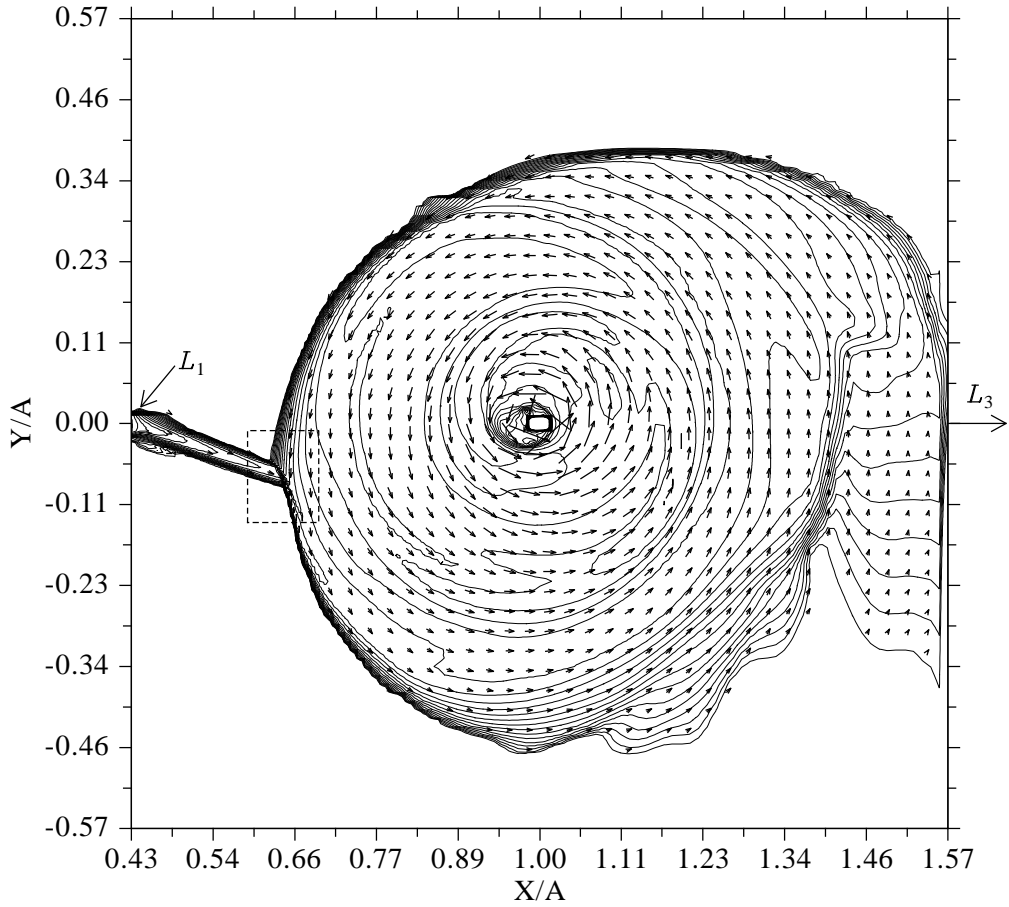,width=13.1cm}}}
\caption{\small Contours of constant density and velocity vectors in
the equatorial plane $XY$ of the system. The shaded rectangle indicates the
zone of interaction between the stream and disk, shown in Figs~\ref{fig7}
and~\ref{fig8}. The
point $\Lone$ and the direction towards $\mathop{\rm L_3}$ are marked.}
\label{fig5}
\end{figure}

\begin{figure}
\centerline{\hbox{\epsfig{figure=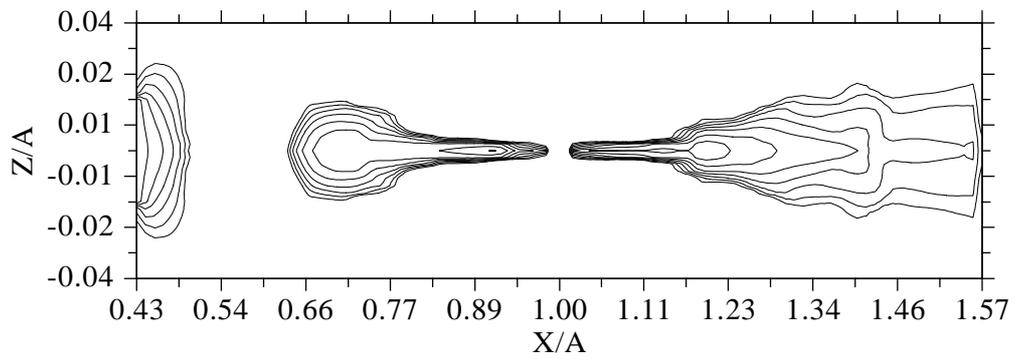,width=13.1cm}}}
\caption{\small Density contours in the frontal plane $XZ$ of the
system.}
\label{fig6a}
\end{figure}
\begin{figure}
\centerline{\hbox{\epsfig{figure=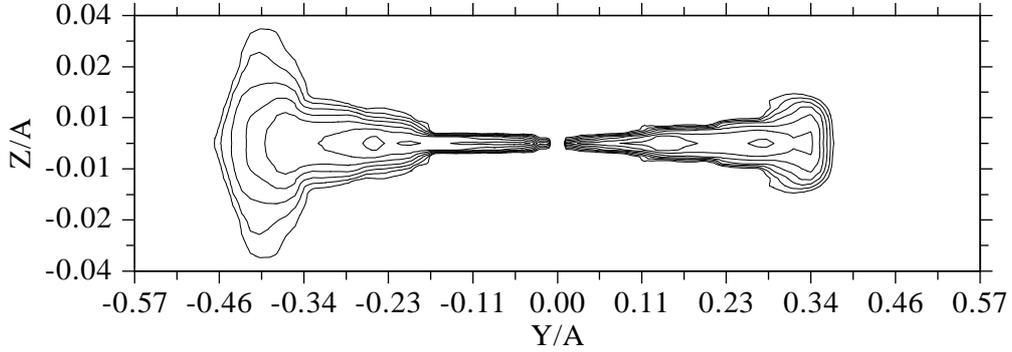,width=13.1cm}}}
\caption{\small Density contours in the plane $YZ$ containing the accretor
and perpendicular to the line connecting the binary components.}
\label{fig6b}
\end{figure}

\begin{figure}
\centerline{\hbox{\epsfig{figure=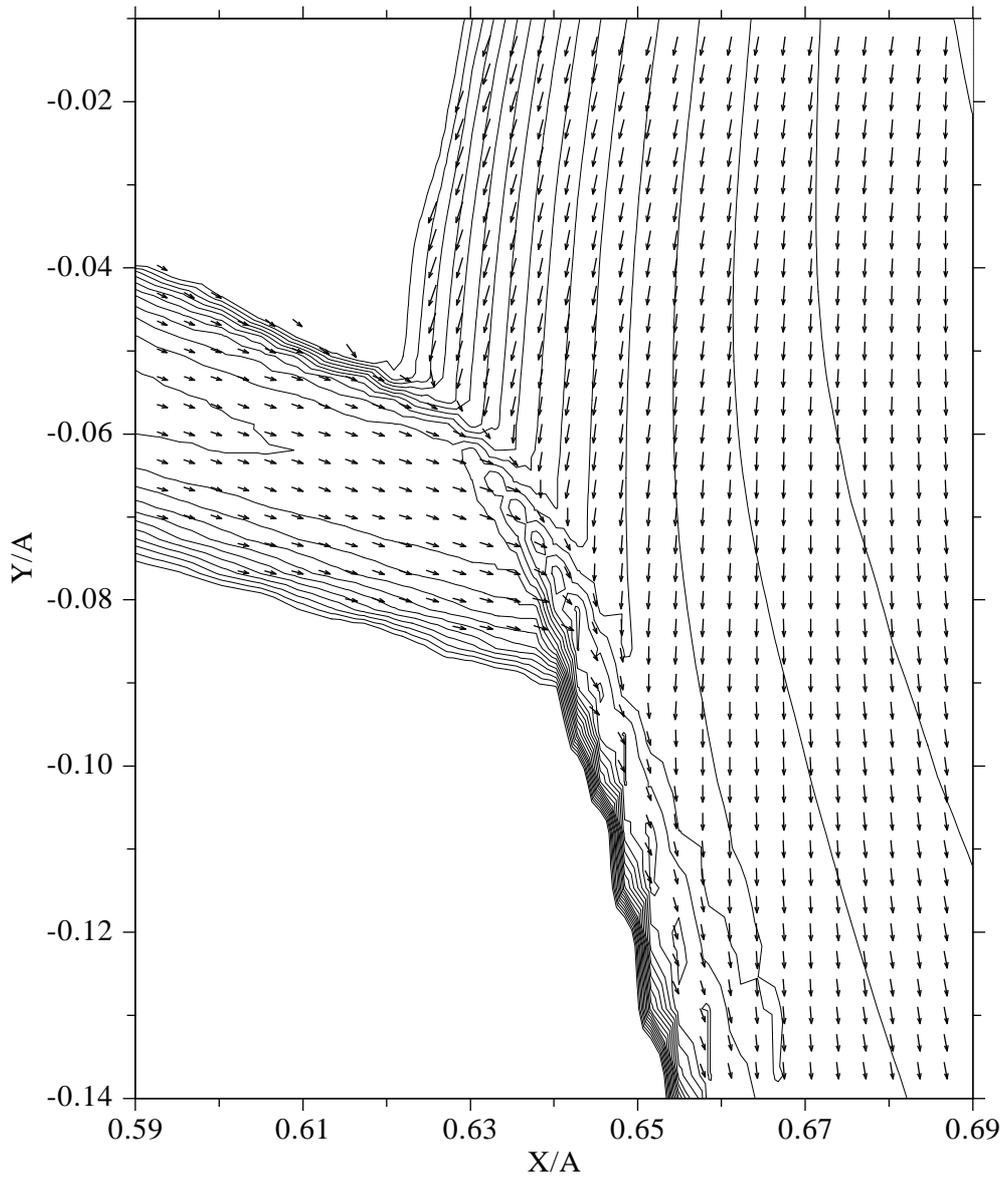,width=13cm}}}   
\caption{\small Contours of constant density and velocity vectors in
the zone of interaction between the stream and disk (the shaded rectangle in
Fig.~\ref{fig5}).}
\label{fig7}
\end{figure}

\begin{figure}
\centerline{\hbox{\epsfig{figure=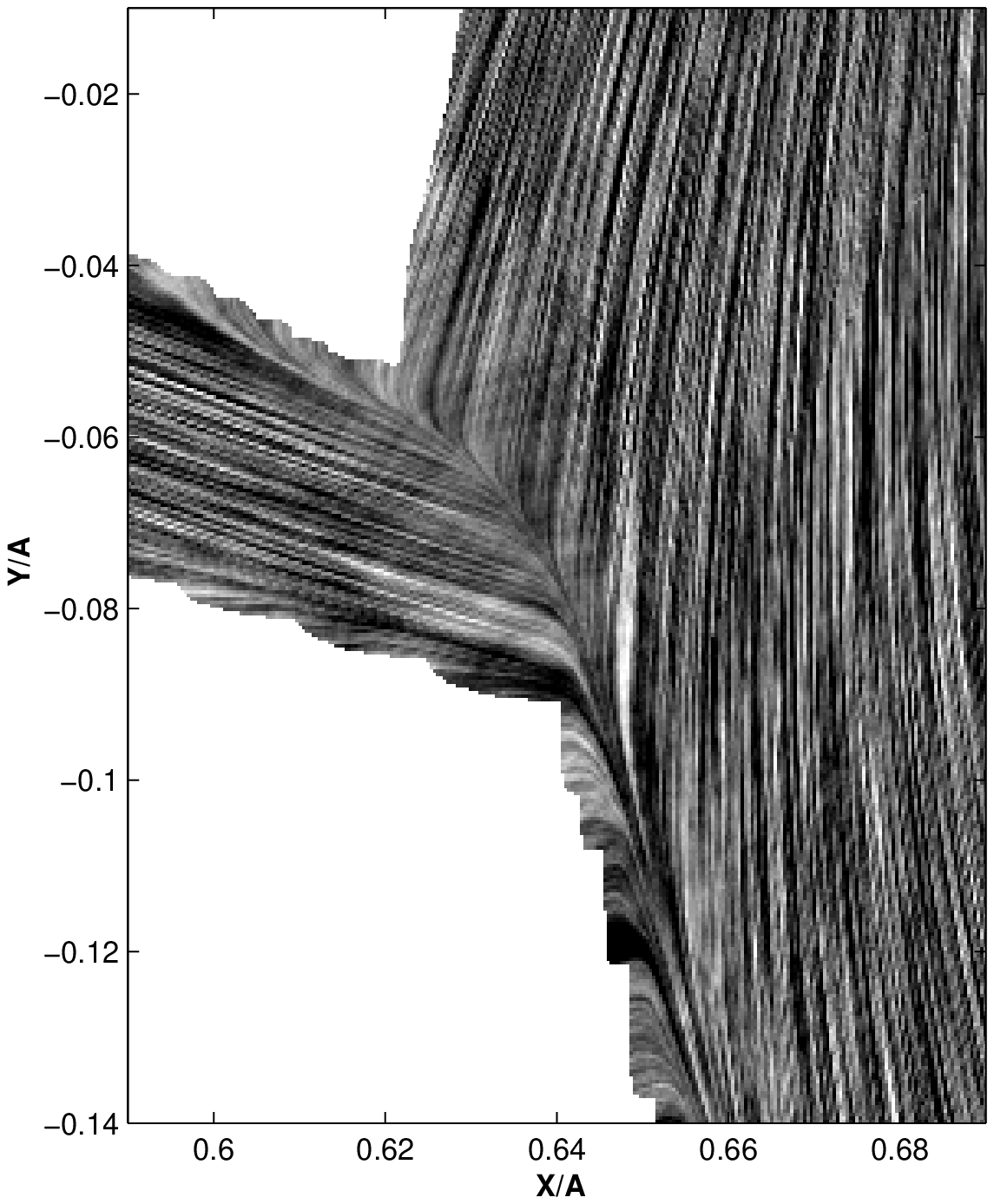,width=13cm}}}
\caption{\small Visualization of the velocity field in the zone of
interaction between the stream and disk (the shaded rectangle in
Fig.~\ref{fig5}).}
\label{fig8}
\end{figure}

Figures~\ref{fig5} to~\ref{fig8} present the morphology of gas flows in
the binary. Figure~\ref{fig5} shows the density and velocity vector
distributions in
the equatorial plane of the system (the $XY$ plane), while Figs.~\ref{fig6a}
and~\ref{fig6b} present density contours in the frontal ($XZ$) plane and in
the $YZ$ plane
containing the accretor and perpendicular to the line connecting the binary
components. In spite of the small height of the forming accretion disk, use
of the JSC parallel-processing computers made it possible to resolve its
vertical structure (the outer parts of the disk were covered by 15 grid
cells, and the inner parts by no fewer than 3 cells). Figure~\ref{fig7}
gives an
enlarged view of the density and velocity vector distributions in the zone of
interaction between the stream and the outer edge of the disk (the area in
the shaded rectangle in Fig.~\ref{fig5}). Figure~\ref{fig8}
presents the so-called texture a
visualization of the velocity field in the zone of interaction between the
stream and disk, constructed using the Line Integral Convolution procedure
(LIC)~\cite{cit30}.

According to our considerations in~\cite{cit8,cit14}, the gasdynamical flow pattern in
a semi-detached binary is formed by the stream from~$\Lone$, the disk, a
circumdisk halo,  and the intercomponent envelope. This subdivision is based
on physical differences between these elements of the flow structure: 
(1) if the motion of the gas is not determined by the gravitational field of
the accretor, it forms the intercomponent envelope; (2) if the gas makes one
revolution around the accretor, but later mixes with the initial stream,
this gas does not become part of the disk, instead forming the circum-disk
halo; (3) the disk is formed by that part of the stream that loses its
momentum and moves towards the center of gravity after entering the
gravitation field of the accretor, rather than interacting with the stream.

In this framework, let us consider the morphology of the flow when the
temperature decreases to 13~600~K over the entire computation domain due to
cooling. Figure~\ref{fig5} indicates that, in this case, the intercomponent envelope
is formed primarily in the vicinity of $\mathop{\rm L_3}$, and does not
affect the solution substantially. We can see from Figs.~\ref{fig5} 
and~\ref{fig6a}--\ref{fig6b} that the
circum-disk halo is pressed against the disk, and its density increases
sharply towards the disk edge.

Figures~\ref{fig7} and~\ref{fig8} show that, in the cool-disk case, the
interaction between
the circum-disk halo and the stream displays all features typical of an
oblique collision of two streams. We can clearly see two shock waves and a
tangential discontinuity between them. The gases forming the halo and stream
pass through the shocks corresponding to their flows, mix, and move along
the tangential discontinuity between the two shocks. Further, this material
forms the disk itself, the halo, and the envelope.

The solution for the cool case displays the same qualitative characteristics
as the solution for the case when the outer parts of the disk are hot: the
interaction between the stream and disk is shockless, a region of enhanced
energy release is formed due to the interaction between the circum-disk halo
and the stream and is located beyond the disk, and the resulting shock is
fairly extended, which is particularly important for explaining the
observations. However, unlike the solution with a high temperature in the
outer regions of the disk~\cite{cit1,cit12,cit14}, in the cool case, the shape of the
zone of shock interaction between the stream and halo is more complex than a
simple ``hot line''.  This is due to the sharp increase of the halo density as
the disk is approached. Those parts of the halo that are far from the disk
have low density, and the shock due to their interaction with the stream
is situated along the edge of the stream. As the halo density increases, the
shock bends, and eventually stretches along the edge of the disk.

\section{Conclusions}
\label{sec5}

Our analysis of the basic processes of heating and cooling in accretion
disks in binaries has shown that, for realistic parameters of the accretion
disks in close binary systems ($\mdot\simeq10^{-12}\div10^{-7}M_\odot/year$
and
$\alpha\simeq10^{-1}\div10^{-2}$), the
gas temperature in the outer parts of the disk is~$10^4$~K to
$\sim10^6$~K.

 Previously, we carried out three-dimensional simulations of the flow
structure in close binaries for the case when the temperature of the outer
parts of the accretion disk was 200--500 thousand~K. Those solutions showed
that the interaction between the stream from the inner Lagrange point and
the disk was shockless. To determine the generality of the solution, the
morphology of the flow for different disk temperatures must be considered.We
have presented here the results of simulations for the case when cooling
decreases the temperature to 13~600~K over the entire computation domain.

Our analysis of the flow pattern for the cool outer parts of the disk confirms
that the interaction between the stream and disk is again shockless. The
computations indicate that the solution for the cool disk case displays the
same qualitative features as in the case when the outer parts of the disk
are hot: the interaction between the stream and disk is shockless, a region
of enhanced energy release formed by the interaction between the circum-disk
halo and the stream is located beyond the disk, and the shock wave that is
formed is fairly extended, and can be considered a ``hot line'' . The cool
solution demonstrates the universal character of our previous conclusions
that the interaction between the stream and disk in semidetached binaries is
shockless.

\section{Acknowledgments}

This work was supported by the Russian Foundation for Basic Research
(project codes 02-02-16088, 02-02-17642), the State Science and Technology
Program in Astronomy, a Presidential Grant of the Russian Federation
(00-15-96722), the Programs of the Presidium of the Russian Academy of
Sciences ``Mathematical Modeling'' and ``Non-steady State Processes
in Astronomy'', and the INTAS Foundation (grant 01-491).

\end{document}